%% file: main.tex
\documentclass[12pt,english]{article}

\input{newcmnds}

\usepackage{fancyhdr}
\allowdisplaybreaks

\begin{document}
\title{Modeling the Dynamics of the COVID-19 Population in Australia: A Probabilistic Analysis}
\author{
	Ali Eshragh\thanks{School of Mathematical and Physical Sciences, University of Newcastle, NSW, Australia, and International Computer Science Institute, Berkeley, CA, USA. Email: \tt{ali.eshragh@newcastle.edu.au}}
	\quad
	Saed Alizamir\thanks{School of Management, Yale University, CT, USA.}
	\quad
	Peter Howley$^\ddagger$ \quad Elizabeth Stojanovski\thanks{School of Mathematical and Physical Sciences, University of Newcastle, NSW, Australia.}
}
\date{}
\maketitle

\vspace*{-0.75cm}
\input{Highlights.tex}

\newpage

\input{Abstract.tex}


\input{Introduction.tex}




\input{DynamicsInAustralia.tex}


\input{Strengths_and_Limitations.tex}


\input{Discussion_and_Conclusion.tex}


\input{Acknowledgment.tex}





\newpage
\input{Appendix.tex}


\end{document}

%% file: newcmnds.tex
\usepackage{algorithm}
\usepackage{booktabs}
\usepackage{graphicx}
\usepackage{amsmath}
\usepackage{amssymb}
\usepackage{latexsym}
\usepackage{crop}
\usepackage{algorithmic,algorithm}
\usepackage{multirow}
\usepackage{bm}
\usepackage{bbm}
\usepackage{enumerate}
\usepackage{url}
\usepackage{array}
\usepackage{paralist}
\usepackage{diagbox}
\usepackage{wasysym}
\usepackage{booktabs}
\usepackage[dvipsnames]{xcolor}
\usepackage[colorlinks = true, pdfstartview = FitV, linkcolor = blue, citecolor = blue, urlcolor = blue]{hyperref}

\usepackage{listings}

\usepackage{rotating}

\usepackage[capitalise]{cleveref}
\crefname{equation}{}{}
\crefname{figure}{Figure}{Figures}
\creflabelformat{equation}{\textup{(#2#1#3)}}
\crefname{assumption}{Assumption}{Assumptions}
\crefname{condition}{Condition}{Conditions}

\usepackage{xspace}

\usepackage{fullpage}
\usepackage{multirow}

\usepackage[sort,nocompress]{cite}

\usepackage{arydshln}
\usepackage{enumitem}
\setlist[enumerate,1]{leftmargin=*,wide=0em, noitemsep,nolistsep, label = {\bfseries \arabic*.}}
\setlist[itemize,1]{leftmargin=*,wide=0em, noitemsep,nolistsep}

\usepackage{pifont}
\makeatletter
\newcommand*{\transpose}{%
	{\mathpalette\@transpose{}}%
}
\newcommand*{\@transpose}[2]{%
	\raisebox{\depth}{$\m@th#1\intercal$}%
}
\makeatother

\renewcommand{\Pr}{\hbox{\bf{Pr}}}

\newcommand*\xbar[1]{%
	\hbox{%
		\vbox{%
			\hrule height 0.5pt 
			\kern0.5ex
			\hbox{%
				\kern-0.1em
				\ensuremath{#1}%
				\kern-0.1em
			}%
		}%
	}%
} 

\definecolor{forestgreen}{rgb}{0.13, 0.55, 0.13}

\definecolor{amber}{rgb}{1.0, 0.75, 0.0}

\definecolor{bananayellow}{rgb}{.8, 0.6, 0}

\newcounter{comment}

\usepackage{amsthm}
\usepackage[framemethod=TikZ]{mdframed}

\mdfdefinestyle{theoremstyle}{%
	linewidth = 1pt,%
	roundcorner = 10pt,%
	leftmargin = 0,%
	rightmargin = 0,%
	backgroundcolor = cyan!3,%
	outerlinecolor = magenta!70!black,%
	splittopskip = \topskip,%
	ntheorem = true,%
}
\mdtheorem[style=theoremstyle]{claim}{Claim}

\newmdtheoremenv[%
linewidth = 1pt,%
roundcorner = 10pt,%
leftmargin = 0,%
rightmargin = 0,%
backgroundcolor = green!3,%
outerlinecolor = blue!70!black,%
splittopskip = \topskip,%
ntheorem = true,%
]{theorem}{Theorem}

\newmdtheoremenv[%
linewidth = 1pt,%
roundcorner = 10pt,%
leftmargin = 0,%
rightmargin = 0,%
backgroundcolor = green!3,%
outerlinecolor = blue!70!black,%
splittopskip = \topskip,%
ntheorem = true,%
]{corollary}{Corollary}

\newmdtheoremenv[%
linewidth = 1pt,%
roundcorner = 10pt,%
leftmargin = 0,%
rightmargin = 0,%
backgroundcolor = green!3,%
outerlinecolor = blue!70!black,%
splittopskip = \topskip,%
ntheorem = true,%
]{lemma}{Lemma}

\newmdtheoremenv[%
linewidth = 1pt,%
roundcorner = 10pt,%
leftmargin = 0,%
rightmargin = 0,%
backgroundcolor = yellow!3,%
outerlinecolor = blue!70!black,%
splittopskip = \topskip,%
ntheorem = true,%
]{definition}{Definition}

\newmdtheoremenv[%
linewidth = 1pt,%
roundcorner = 10pt,%
leftmargin = 0,%
rightmargin = 0,%
backgroundcolor = green!3,%
outerlinecolor = blue!70!black,%
splittopskip = \topskip,%
ntheorem = true,%
]{proposition}{Proposition}

\newmdtheoremenv[%
linewidth = 1pt,%
roundcorner = 10pt,%
leftmargin = 0,%
rightmargin = 0,%
backgroundcolor = green!3,%
outerlinecolor = blue!70!black,%
splittopskip = \topskip,%
ntheorem = true,%
]{condition}{Condition}

\newmdtheoremenv[%
linewidth = 1pt,%
roundcorner = 10pt,%
leftmargin = 0,%
rightmargin = 0,%
backgroundcolor = green!3,%
outerlinecolor = blue!70!black,%
splittopskip = \topskip,%
ntheorem = true,%
]{assumption}{Assumption}

\theoremstyle{definition}
\newmdtheoremenv[%
linewidth = 1pt,%
roundcorner = 10pt,%
leftmargin = 0,%
rightmargin = 0,%
backgroundcolor = blue!3,%
outerlinecolor = blue!70!black,%
splittopskip = \topskip,%
ntheorem = true,%
]{example}{Example}

\theoremstyle{definition}
\newmdtheoremenv[%
linewidth = 1pt,%
roundcorner = 10pt,%
leftmargin = 0,%
rightmargin = 0,%
backgroundcolor = red!3,%
outerlinecolor = blue!70!black,%
splittopskip = \topskip,%
ntheorem = true,%
]{remark}{Remark}

\usepackage{tikz}
\usepackage{xparse}

\NewDocumentCommand\DownArrow{O{2.0ex} O{black}}{%
	\mathrel{\tikz[baseline] \draw [<-, line width=0.5pt, #2] (0,0) -- ++(0,#1);}
}

\usepackage{listings} 

\definecolor{mygreen}{rgb}{0,0.6,0}
\definecolor{mygray}{rgb}{0.5,0.5,0.5}
\definecolor{mymauve}{rgb}{0.58,0,0.82}

\newcommand {\Ex} { {\mathbb E} }

\usepackage{dsfont}

\usepackage[latin1]{inputenc}
\usepackage{amsmath}
\usepackage{amsfonts}
\usepackage{amssymb}
\usepackage{makeidx}
\usepackage{graphicx}
\usepackage{caption}
\usepackage{subcaption}
\usepackage{framed}
\usepackage{booktabs,array}
\usepackage{xcolor}

\newcommand{\superscript}[1]{\ensuremath{^{\textrm{#1}}}}
\usepackage{scrextend}

%% file: Highlights.tex
{\small
	
\section*{Highlights}

\begin{itemize}
	\item \medskip This work applies a novel and effective approach using a partially-observable stochastic process to study the dynamics of the COVID-19 population in Australia over the $1$ March--$22$ May $2020$ period.

	\item \medskip The key contributions of this work include (but are not limited to):
	
	\begin{enumerate}[label = (\roman*), leftmargin = 1.2\parindent]
		\item \medskip identifying two structural break points in the numbers of new cases coinciding with where the dynamics of the COVID-19 population are altered: the \emph{first}, a major break point, on $27$ March $2020$, is one week after implementing the ``lockdown restrictions'', and the \emph{second} minor point on $18$ April $2020$, is one week after the ``Easter break'';
		
		\item \medskip forecasting the future daily numbers of new cases up to $28$ days in advance with extremely low mean absolute percentage errors (\textsf{MAPE}s) using a relative paucity of data, namely, \textsf{MAPE} of $1.53\%$ using $20$ days of data to predict the number of new cases for the following $6$ days, \textsf{MAPE} of $0.43\%$ using $34$ days of data to predict the number of new cases for the following $14$ days, and \textsf{MAPE} of $0.20\%$ using $55$ days of data to predict the number of new cases for the following $28$ days;
				
		\item \medskip estimating approximately $33\%$ of COVID-19 cases as unobserved by $26$ March $2020$, reducing to less than $5\%$ after implementing the Government's constructive restrictions;
		
		\item \medskip predicting that the growth rate, prior to the Government's implementation of restrictions, was on a trajectory to infect numbers equal to Australia's entire population by $24$ April $2020$;

		\item \medskip \label{item:contributions} estimating the dynamics of the growth rate of the COVID-19 population to slow down to a rate of $0.820$ after the first break point, with a slight rise to $0.979$ after the second break point;

		\item \medskip Advocating the outlined stochastic model as practically beneficial for policy makers when considering implementation and easing of virus restrictions due to the demonstrated sensitivity of the dynamics of the COVID-19 population in Australia to both major and minor system changes. 
	\end{enumerate}

	\item \medskip The model developed in this work may further assist policy makers to consider the impact of several potential scenarios in their decision-making processes.
\end{itemize}}

%% file: Abstract.tex
\begin{abstract}
	The novel Corona Virus COVID-19 arrived on Australian shores around $25$ January $2020$. This paper presents a novel method of dynamically modeling and forecasting the COVID-19 pandemic in Australia with a high degree of accuracy and in a timely manner using limited data; a valuable resource that can be used to guide government decision-making on societal restrictions on a daily and/or weekly basis. The ``partially-observable stochastic process'' used in this study predicts not only the future actual values with extremely low error, but also the percentage of unobserved COVID-19 cases in the population. The model can further assist policy makers to assess the effectiveness of several possible alternative scenarios in their decision-making processes.
\end{abstract}

%% file: Introduction.tex
\section{Introduction: COVID-19 Pandemic}

The novel beta-coronavirus, later named ``COVID-19'', was first reported in late December $2019$ in Wuhan City, China \cite{Huang20}. Early reports indicated a wet market in Wuhan to be the origin of the outbreak, affecting approximately $66\%$ of market staff, and comprising symptoms resembling pneumonia of fever, dry cough, and fatigue \cite{Wu20}. The market closed $1$ January $2020$, following an epidemiologic alert announced by the local health authority in China on $31$ December $2019$. The infection was reported to have spread to many cities across China over January $2020$, with thousands in China becoming infected by the disease, while also spreading rapidly globally, affecting countries including Thailand, Japan, Korea, Vietnam, Singapore, United States and Germany \cite{Wu20}. The World Health Organization (WHO) declared the outbreak a pandemic on $11$ March $2020$ and, as of $22$ May $2020$, a total of $4,993,470$ confirmed cases of COVID-19 globally were reported by WHO with $327,738$ related deaths across at least $216$ countries\footnote{\label{ft:WHO_reports}\url{https://www.who.int/emergencies/diseases/novel-coronavirus-2019/situation-reports/}}.

\paragraph{Pandemic in Australia. }According to official reports, the novel Corona Virus COVID- 19 arrived on Australian shores on around $25$ January $2020$. From $5$ March $2020$, the number of new cases grew rapidly and reached over $300$ cases daily in late March\footref{ft:WHO_reports}. Following lockdown restrictions by the Australian Government from mid-March, the daily number of new cases started declining from early April, reaching approximately $20$ cases daily by late April\footref{ft:WHO_reports}.

Preventative measures to minimize transmission were increasingly imposed by the Australian Government from $1$ February $2020$ with foreign nationals from mainland China banned entry to Australia, and $14$ days of self-quarantining imposed for returning citizens from China\footnote{\url{https://www.pm.gov.au/media/extension-travel-ban-protect-australians-coronavirus}}. Travel restrictions were subsequently imposed with all travelers arriving to Australia required to self-isolate for $14$ days from $15$ March $2020$, with fines of up to AUD$\$50,000$ for non-compliance\footnote{\label{ft:BBC_report}\url{https://www.bbc.com/news/world-australia-51894322}}. A general travel ban was imposed from $20$ March $2020$ with Australia closing its borders to all non-residents\footnote{\url{https://www.australia.gov.au/coronavirus-updates}}. 

A human bio-security emergency was declared in Australia on $18$ March $2020$ with a social distancing rule of $4$ square meters per person imposed from $20$ March $2020$. From $22$ March $2020$, a mandatory closure of non-essential services was imposed with some states closing their borders allowing only the state's residents to return\footref{ft:BBC_report} and from $23$ March $2020$, all places of social gathering were closed with cafes and restaurants limited to takeaway\footnote{\label{ft:GovCOVID_updates}\url{https://www.australia.gov.au/coronavirus-updates}}. From $29$ March $2020$, public gatherings were limited to two people if they were not from the same household and there were only four acceptable reasons for leaving homes, comprising shopping for essentials, medical or compassionate needs, exercise and work or education purposes\footref{ft:GovCOVID_updates}.

Some subtle variation in the timing of the implementation of these measures occurred between States/Territories with State Governments/Territory officials also imposing additional restrictions in response to State/Territory-specific data. For example, some states introduced social distancing measures in schools from $15$ March $2020$, preventing students and staff from congregating in large numbers with several university graduations, conferences, events, classes and student organized events also canceled\footnote{\url{https://www.smh.com.au/national/nsw/school-assemblies-excursions-and-events-to-be-cancelled-20200315-p54aae.html}}.

At the time of writing, the Government had a three-stage plan to reopen Australia by July $2020$\footnote{\url{https://www.pm.gov.au/media/update-coronavirus-measures-08may20}}. The three stages reflect increasing the numbers of permissible visitors in homes and public places, whilst still maintaining noted hygiene and social spacing, along with the opening of various places of employment and social interaction (restaurants, community centers in Stage $1$). Accompanying this are the lifting of travel restrictions: local and regional in Stage $1$, interstate in Stage $2$ and partial international, principally Pacific region, in Stage $3$. The seven States and Territories invoke these at slightly differing times to reflect their local experiences and numbers of infected people.     

\paragraph{Vital need for modeling. }The outbreak of COVID-19 and its accompanying pandemic has created an unprecedented challenge and unilateral response worldwide, and urged every nation to deploy its utmost resources toward combating the disease whilst managing the economic and social impacts. Tracking the epidemic and estimating the size of the infected population and effects of potential guidelines and restrictions has become a critical priority for most governments around the globe as it has immediate ramifications on all subsequent policy interventions (e.g. see \cite{Che20,Dandekar20,KOO20,Moss20,Small20}).

Stochastic processes are designed to deal with change that involves randomness and uncertainty, both aspects that are paramount to the COVID-19 outbreak. In particular, \emph{partially-observable stochastic processes} specifically account for incomplete knowledge of a system that arises from knowing only partially about a given situation, without knowledge of the complete situation. A common form of partial observation is one whereby the state of the system, or each component of the system, can be observed with only a certain degree of certainty. An example is the case of biological invasions, whereby an invasive species or individual of the species can be detected with only a certain probability upon each survey (e.g. see \cite{OBKP12}). Another application is in medical testing, where a test can provide a false negative, so that the infection can then only be detected with a certain probability for an infected individual upon administering the test (e.g. see \cite{Kao03,Ger09}). Further applications of partially-observable stochastic processes include (but are not limited to) recognizing patterns \cite{Fin89}, analyzing digital signals \cite{Vas06}, and understanding biological processes \cite{AE98,BB01}.

In this paper, we focus on modeling the early stages of the COVID-19 outbreak in Australia, and provide an epidemic model that complements others in use by providing extremely accurate estimates of COVID-19 transmission in Australia, including estimates of hidden cases, in timeframes relevant to policy implementation. More precisely, we utilize a special class of stochastic processes, the \emph{partially-observable pure birth process}, to model the dynamics of the COVID-19 population in Australia. In the present epidemiological context of modeling the COVID-19 outbreak, the main source of uncertainty comes from the stochastic dynamics of the system as well as the structure of sampling in which each infected individual can be tested with only a certain degree of certainty. Our model particularly suits situations where the number of infected citizens is relatively minimal compared to the total at-risk population, which is the case for the majority of regional and national jurisdictions. This is a critical phase of disease spread, and requires policy measures that effectively control growth. The effectiveness of these policies, in turn, depends heavily on the quality of the models used and the precision of the estimates that they generate. The following two features of our model establish its benefits relative to other models (such as Susceptible, Infected and Recovered (\texttt{SIR}) model):
\begin{itemize}[leftmargin = 0.9\parindent]
	\item \medskip The \emph{robust predictive nature} where, with only small amounts of data, the (subsequently released) future actual values are forecasted very well;
		
	\item \medskip The capability to estimate not only the growth rate of the COVID-19 cases, but also the percentage of \emph{unobserved cases}, which represent those in the population who have not been officially diagnosed.
\end{itemize}
\medskip 

The Highlights identify the main contributions of this work. In summary, the novel modeling identifies key break points associated with social restrictions imposed by the Government; estimates the percentage of unobserved cases; accurately predicts future numbers of COVID-19 cases pre- and post-implementation of restrictions; demonstrates how growth rates in cases changes in response to major and minor break points; and provides guidance for policy makers in terms of the sensitivity of the dynamics of COVID-19 in Australia.

The structure of this paper is as follows: \cref{sec:covid_in_australia} applies a partially-observable pure birth process to model the dynamics of the COVID-19 population in Australia and predicts future values as well as the percentage of unobserved hidden cases. \cref{sec:Strengths_and_Limitations} discusses the strengths and limitations of the model used in this study. \cref{sec:Discussion_and_Conclusion} provides the final discussions on policy implications and concluding remarks. \cref{sec:Appendix} presents an overview of partially-observable continuous-time Markov population processes along with their theoretical and applied properties. 

%% file: DynamicsInAustralia.tex
\section{Modeling: Dynamics of the COVID-19 Population in Australia} \label{sec:covid_in_australia}

A \emph{Continuous-time Markov population Process} (\texttt{CTMPP}) is a class of stochastic processes often used to model biological phenomena (e.g. see \cite{Ren91,KR08,BM12}). The study of a \texttt{CTMPP} under partial observations, referred to as a ``partially-observable continuous-time Markov population process'' (\texttt{PO-CTMPP}) is of interest for the present study. A special class of \texttt{PO-CTMPP}s is the \emph{partially-observable pure birth process} (\texttt{PO-PBP}) whereby, while the underlying model is a stochastic ``pure birth process'', observations are made partially according to a binomial distribution. 

Bean et al. \cite{BEER_POCMP} extensively studied the theoretical properties of a \texttt{PO-CTMPP} and a \texttt{PO-PBP}, and derived the conditional probability distribution of the true state of the system and future values of partial observations, given the history of partial observations. Furthermore, they showed that, unlike a pure birth process, a \texttt{PO-PBP} is not Markovian of any order. Bean et al. \cite{BER_FI} applied these results to find the Fisher Information for a \texttt{PO-PBP} and derived the optimal experimental design. The details of these results are summarized in \cref{sec:Appendix}. 

We utilize these approaches here to model and analyze the dynamics of the COVID-19 population in Australia. Due to practical limitations described in \cref{sec:POCTMPM}, not all infected cases may be observed each day, implying that the confirmed cases reported officially are only partial representations of actual cases. Therefore, a \texttt{PO-PBP} can be considered a superior model to explain the complex dynamics of the COVID-19 population.

\begin{remark}
	We assume that there is no shortage of ``COVID-19 test kits'' in Australia. If this assumption were false, the model would lose applicability once the required sampling rate reached the limit imposed by kit shortages. However, this has not happened to date in Australia, according to the Australian Prime Minister's statements during the pandemic\footnote{\url{https://www.linkedin.com/in/scottmorrisonmp/}}. 
\end{remark} 

The data for this study are obtained from ``daily WHO reports''\footnote{It should be noted that there are a few official resources for the COVID-19 data with minor differences in their reports. Although the structure of our model and main output stay consistent for the data from different resources, the numerical results might slightly vary.} provided on their website\footnote{\url{https://www.who.int/emergencies/diseases/novel-coronavirus-2019/situation-reports/}}. All algorithms are coded in \textsc{C}, and the outputs are analyzed in \textsc{Matlab} R2019a.

\begin{figure}[h!]
	\centering
	\includegraphics[width=12cm]{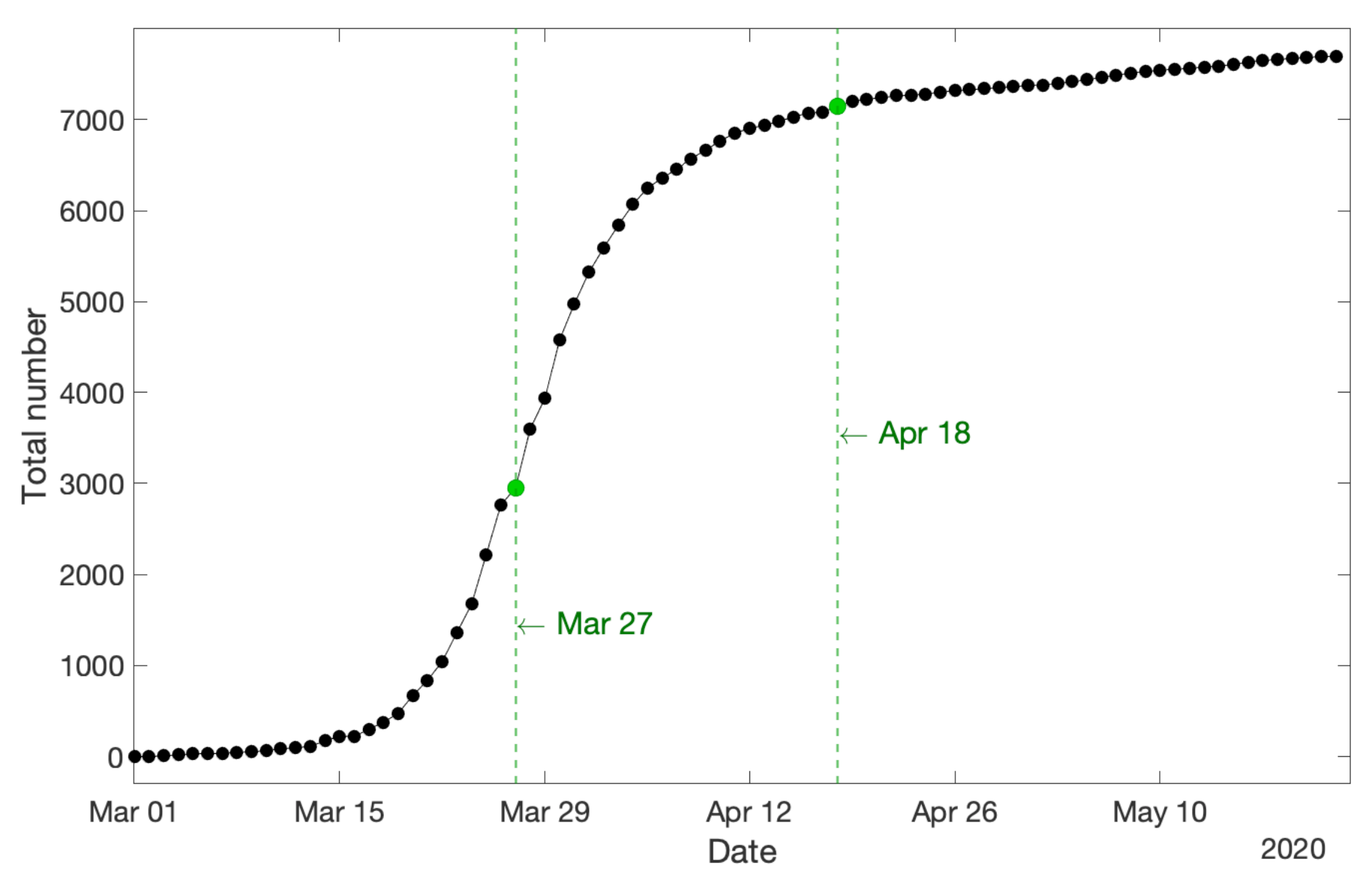}
	\caption{Cumulative new COVID-19 cases in Australia for the whole span of the data. Two structural break points in the dynamics of the COVID-19 population are evident: a major break point on $27$ March $2020$, and a minor break point on $18$ April $2020$.}
	\label{fig:cumsum_Aust}
\end{figure}

The first step in data analysis is visualization. \cref{fig:cumsum_Aust} displays the cumulative new COVID-19 cases in Australia from $1$ March to $22$ May $2020$ and demonstrates two \emph{structural break points} where the population dynamics have been altered, mainly attributed to new polices and exogenous factors: 
\begin{enumerate}[label = (\roman*)]
	\item \medskip The first break point occurs on $27$ March $2020$. This point corresponds to one week after implementation of the ``lockdown restrictions''. As shown in \cref{fig:cumsum_Aust}, this is a crucial break point where the curvature of the cumulative new cases dramatically changes from a convex exponential growth to a concave stable pattern. Furthermore, it is observed that the growth rate starts declining after this break point.
	
	\item \medskip The second break point occurs on $18$ April $2020$ which corresponds to a week after the ``Easter break'' in Australia. Unlike the first break point, the second one does not transform the curvature or stability of the graph, but instead shifts it up slightly and slows down the speed at which the growth rate parameter is declining.  
\end{enumerate}
\medskip

\begin{remark}
		\cref{fig:cumsum_Aust} along with those two structural break points indicate that the dynamics of the COVID-19 population in Australia appear very sensitive towards major/minor system changes, which should be a serious consideration for policy makers while easing out virus restrictions.
\end{remark}

To gain insights into the complex nature of the population for modeling its dynamics, we carry out our analysis in three nested real time steps:  
\begin{itemize}
	\item \medskip Step 1: $1-26$ March $2020$, until the first break point (cf. \cref{sec:PartI}); 
	
	\item \medskip Step 2: $1$ March--$17$ April $2020$, until the second break point (cf. \cref{sec:PartII});
	 
	\item \medskip Step 3: $1$ March--$22$ May $2020$, the whole span of the data (cf. \cref{sec:PartIII}). 
\end{itemize}
\medskip

The general scope of our modeling for each step is as follows: we fit a \texttt{PO-PBP} to the data to model the dynamics of the COVID-19 population over the designated period. A \texttt{PO-PBP} possesses two major parameters comprising the growth rate $\lambda_t$ and the observation probability $p_t$ (cf. \ref{sec:POPBP}). We construct the likelihood function of partial observations according to \cref{eq:LF-chain_rule} in conjunction with \cref{cor:PO-PBP}, and truncate the involved infinite sums by utilizing \cref{eq:truncation} and \cref{pro:Ex_Var}. Then, the logarithm of the likelihood function is maximized over the range of parameters to find their maximum likelihood estimates (\textsf{MLE}s). Finally, \cref{pro:Ex_Yt} is applied to predict the future values of partial observations. 

Furthermore, in order to evaluate the accuracy of predictions generated by the estimated models, the dataset for each nested real time step is partitioned into two mutually exclusive segments consisting of the \emph{training data} to estimate the parameters and forecast future values, and the \emph{test data} to evaluate the accuracy of predictions. To measure the latter, we utilize \emph{mean absolute percentage error}, introduced in \cref{def:MAPE}. 

\begin{definition}[\cite{Hyndman06}]\label{def:MAPE}
	The mean absolute percentage error (\textsf{MAPE}) is defined as:
	\begin{align*}
		\mathsf{MAPE} & = \frac{1}{h}\sum_{t=1}^{h}\left\lvert\frac{  f_t-x_t }{x_t}\right\rvert\times 100\%,
	\end{align*}	
	where  $f_t$, $x_t$ and $h$ are the forecasted values, actual values, and prediction horizon, respectively. 
\end{definition}

\begin{remark}
	It should be noted that the quality of predictions reported in \cref{sec:PartI,sec:PartII,sec:PartIII} is robust to the choice of training and test data, provided there are enough observations in the former set (cf. \cref{fig:sensitivity}).  
\end{remark}

\subsection{Step 1: 1--26 March 2020}\label{sec:PartI}

This step involves the beginning of the pandemic in Australia where the COVID-19 population is growing exponentially fast. The data from $1-20$ March $2020$ are used as the {\emph{training data}} to estimate the parameters of the model, and the data from $21-26$ March $2020$ are used as the \emph{test data} to evaluate the accuracy of predictions. For this period of modeling, we consider the flowing dynamics for the two parameters growth rate $\lambda_t$ and observation probability $p_t$ for the underlying \texttt{PO-PBP}: 
\begin{align*}
	\lambda_t & = \alpha_1^{(t - t_0)} \lambda & \mbox{for $t$ from $1-26$ March }2020,
\end{align*}
and
\begin{align*}
	p_t & = \min\{\beta_1^{(t - t_0)} p, 1\} & \mbox{for $t$ from $1-26$ March }2020,
\end{align*}
where, $\alpha_1 > 0$ and $\beta_1 > 0$ are constant coefficients, $\lambda$ and $p$ are unknown initial values of the parameters, and $t_0$ is the date of the first observation (i.e., $1$ March $2020$). After constructing the likelihood function and maximizing over the parameters, the \textsf{MLE}s in \cref{tab:MLE_lambda_p-PartI} are derived. Since \textsf{MLE} of $\alpha_1$ and $\beta_1$ turn out equal to $1$, both $\mathsf{MLE}(\lambda_t) = 0.235$ and $\mathsf{MLE}(p_t) = 0.67$ are fixed for all $t$ in the range.
\begin{table}[h!]
\begin{center}
	\begin{tabular}{ | l | c | c | c | c |}
		\hline
		\textbf{Parameters} & $\lambda$ & $\alpha_1$ & $p$ & $\beta_1$  \\ \hline
		\textbf{\textsf{MLE}} & 0.235 & 1.000 & 0.67 & 1.00  \\ \hline
	\end{tabular}
	\caption{\textsf{MLE} of parameters}
	\label{tab:MLE_lambda_p-PartI}
\end{center}
\end{table}

\begin{figure}[h!]
	\centering
	\includegraphics[width=12cm]{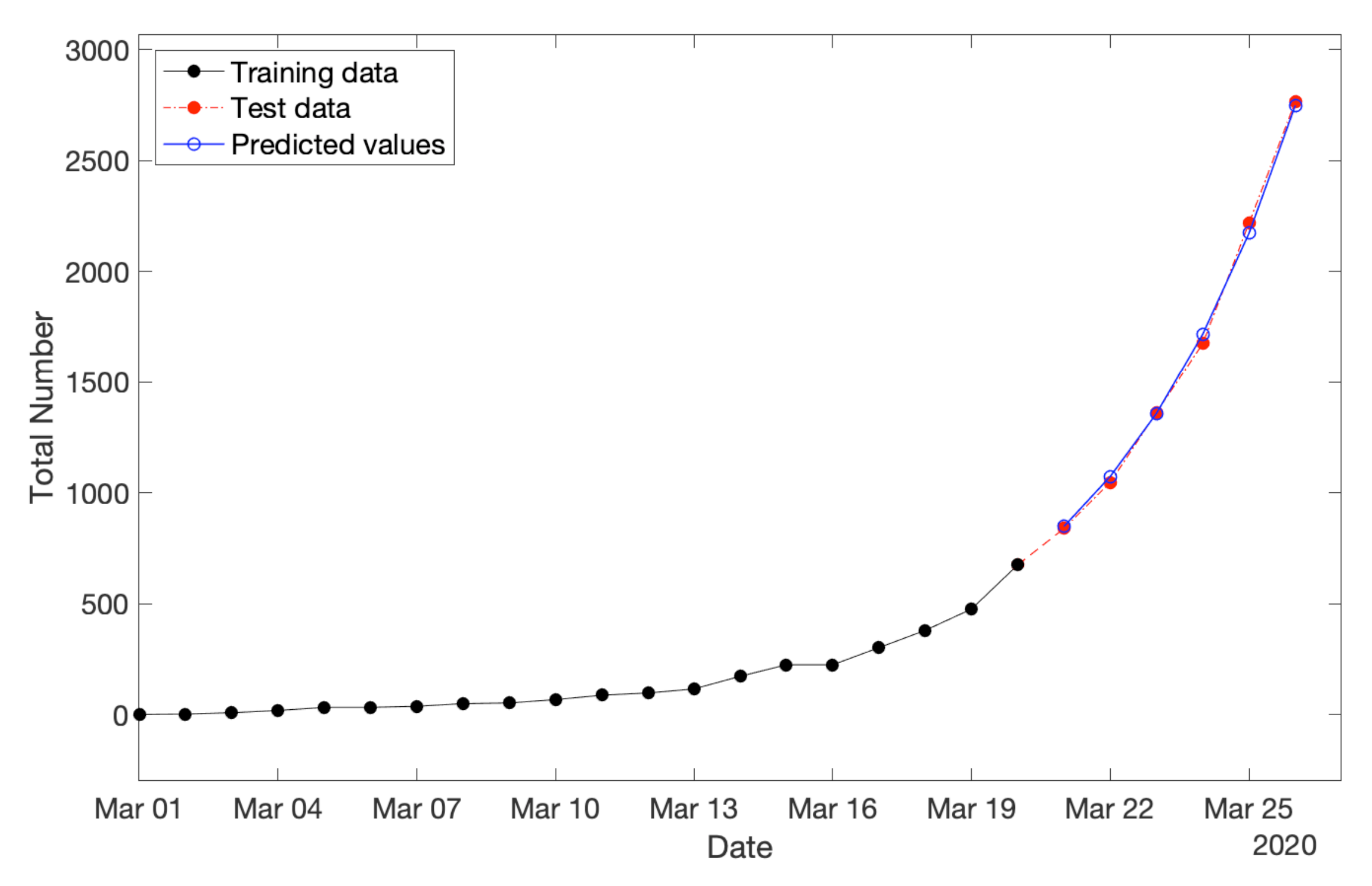}
	\caption{Cumulative new COVID-19 cases in Australia for three categories consisting of: (i) training data spanning from $1-20$ March $2020$ (in black), test data spanning from $21-26$ March $2020$ (in red), and (iii) predicted values over the period of test data (in blue) with $\mathsf{MAPE} = 1.53\%$.}
	\label{fig:PartI_1}
\end{figure}

By applying \cref{pro:Ex_Yt}, the expected values of partial observations over the span of test data (i.e., $21-26$ March $2020$) are predicted. The results are shown in \cref{fig:PartI_1}, where the training data, test data, and predictions are displayed by the black solid plot, red dot-dash plot, and blue solid plot, respectively. It is readily seen that the predicted values are so well fitted to the actual test data. This observation is numerically confirmed with $\mathsf{MAPE} = 1.53\%$. 

\begin{figure}[h!]
	\centering
	\includegraphics[width=12cm]{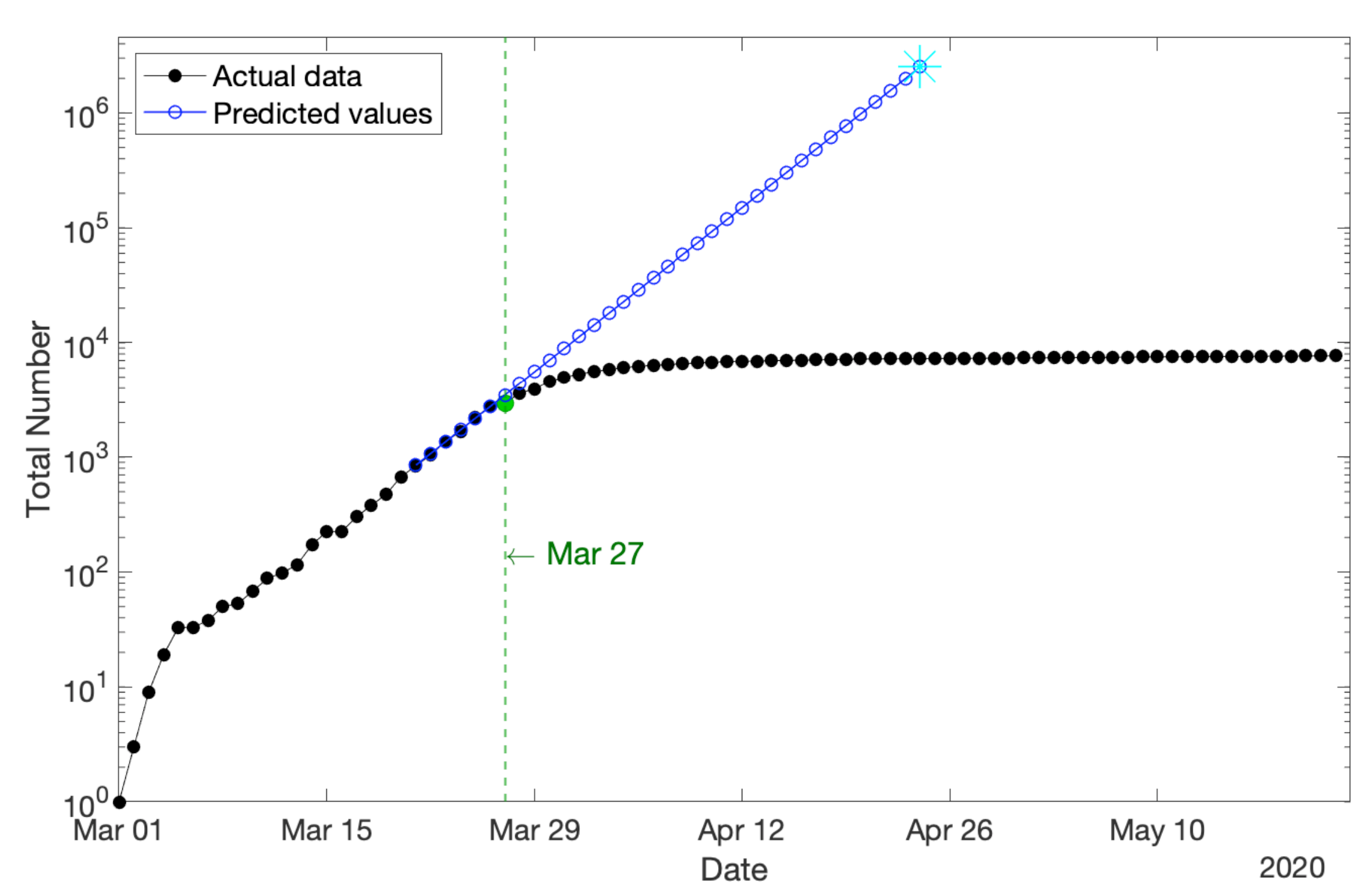}
	\caption{Cumulative new COVID-19 cases in Australia for the whole span of the data (in logarithmic scale) compared with the predicted values for the scenario in which the lockdown restrictions had not been implemented. It shows the exponential growth of the COVID-19 population before the impact of lockdown restrictions on $27$ March $2020$ (marked in green).}
	\label{fig:PartI_2}
\end{figure}

The model additionally suggests that prior to the government's implementation of restrictions, the growth rate was on a trajectory to hit infection numbers equal to Australia's entire population by $24$ April $2020$, a prediction which would have probably been softened only somewhat by limiting factors such as our island status. This asserts the effectiveness of Government's policies and restrictions. \cref{fig:PartI_2} displays the semi-log plot (where the $y$-axis is scaled logarithmically) of the cumulative new COVID-19 cases in Australia (from $1$ March--$22$ May $2020$) and the model predictions (from $21$ March--$24$ April $2020$) in black and blue, respectively. This figure reveals the exponential growth of the COVID-19 population before the impact of lockdown restrictions on $27$ March $2020$ (marked in green). 

\begin{remark}
	\cref{fig:PartI_2} shows an initial break point on $6$ March $2020$. As it is located at the outset of the pandemic in Australia and the number of confirmed cases is still very small during that short period, we disregard it as a break point in our analysis. It does not have a significant influence on the results. 
\end{remark}

Finally, the \textsf{MLE} of the observation probability $p_t$ estimates that only $67\%$ of COVID-19 cases in Australia had been tested by $26$ March $2020$, and the hidden $33\%$ cases had not been recorded/diagnosed officially, by this date.  

\paragraph{Identifiability analysis. }In statistical inference, there are several tools to measure the quality of estimates, including \emph{identifiability}. 
\begin{definition}[\cite{Lehmann}]
	A statistical inference is called ``identifiable'', if different values of the model parameters generate different probability distributions of the observable variables.
\end{definition} 
If an inference is truly not identifiable, then mathematically, the value of the likelihood function will be a constant at all values of the parameters which are equivalent. Therefore, in this case, one would expect to see a ridge on the likelihood surface of roughly constant values as the parameters change.  

\begin{figure}[h!]
	\centering
	\includegraphics[width=12cm]{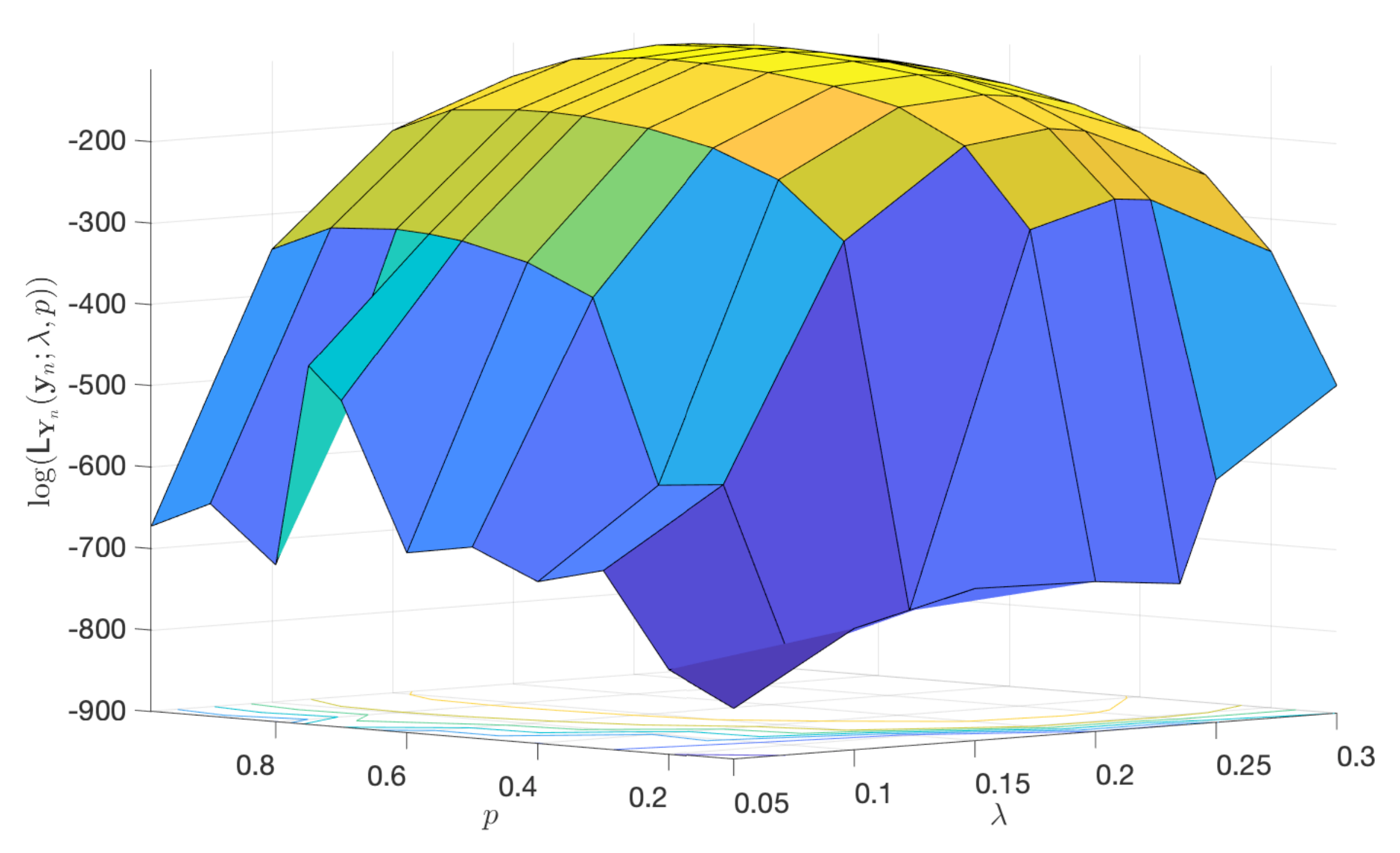}
	\caption{The log-likelihood function of partial observations in terms of the two parameters $\lambda$ and $p$. This plot illustrates the identifiability of the estimates provided in \cref{tab:MLE_lambda_p-PartI}.}
	\label{fig:identifiability}
\end{figure}

In order to see the identifiability of the \textsf{MLE}s of the main parameters $\lambda$ and $p$ given in \cref{tab:MLE_lambda_p-PartI}, we plot the log-likelihood function of partial observations in terms of the two parameters. As depicted in \cref{fig:identifiability}, it is clearly observed that there exists a curvature in the log-likelihood function, illustrating these estimates are identifiable.  

\begin{remark}\label{rem:rule_of_thumb}
There is a heuristic that so long as the log-likelihood function changes by at least $2$ units, then it is regarded as a worthwhile change, implying the identifiability of estimates. So, by considering the locus of points in $(\lambda,p)$ that remain within $2$ units of the log-likelihood function at $(\mathsf{MLE}(\lambda)=0.235,\mathsf{MLE}(p)=0.67)$, the locus allows for $\mathsf{MLE}(p)$ to range within $[0.55,0.75]$ and for $\mathsf{MLE}(\lambda)$ to range within $[0.225,0.245]$. A few other points within those two ranges are chosen as the \textsf{MLE}s of $\lambda$ and $p$, but no significant difference in \textsf{MAPE} of predictions is observed. 
\end{remark}

\begin{figure}[h!]
	\centering
	\includegraphics[width=12cm]{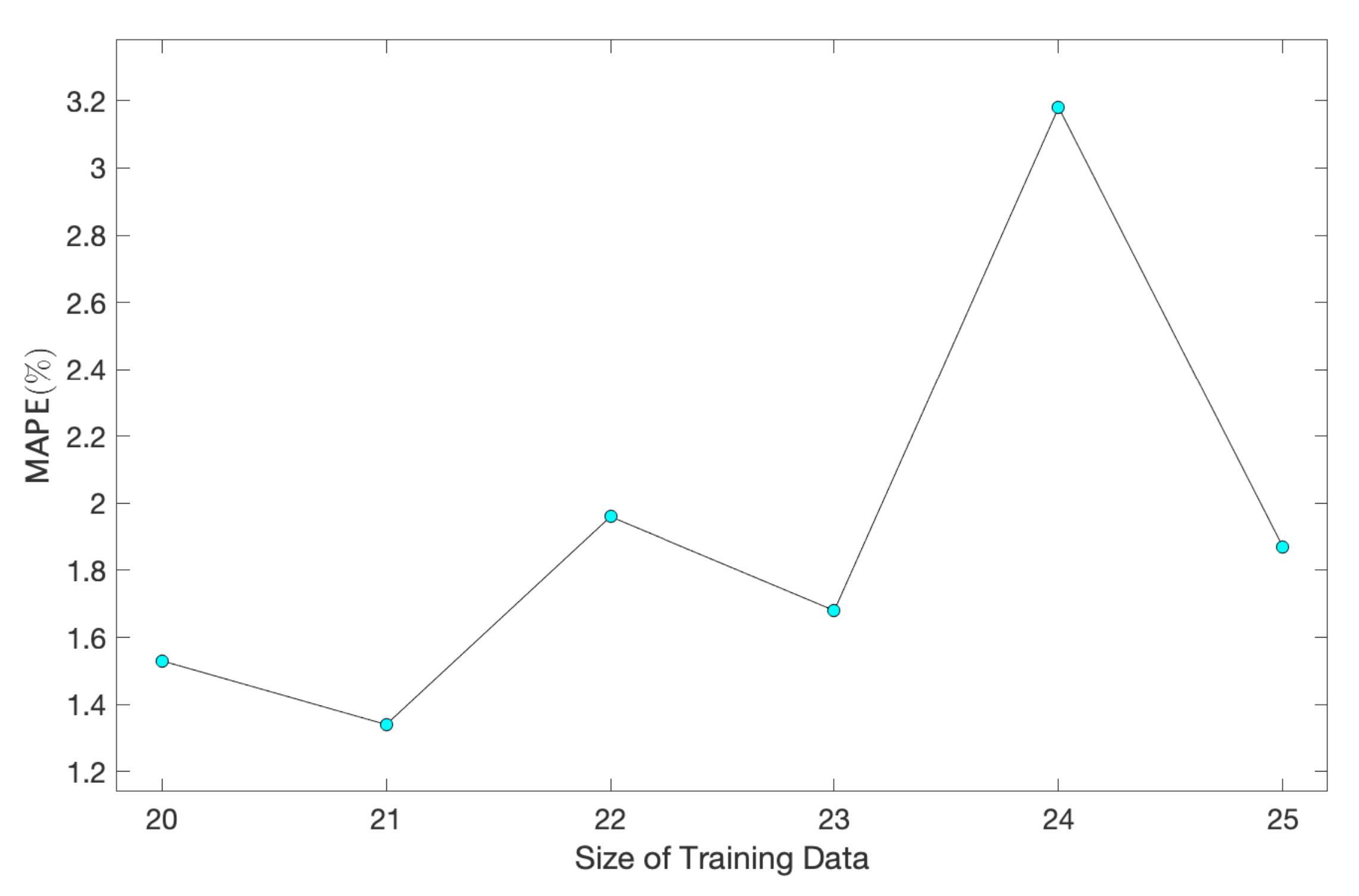}
	\caption{\textsf{MAPE} of predictions vs. size of training data for a range of size of training data (test data) varying from $20-25$ ($6-1$).}
	\label{fig:sensitivity}
\end{figure}

\paragraph{Sensitivity analysis. }Due to the very small amount of available data ($26$ observations in this step, and $76$ data in total), there is limited opportunity to investigate the robustness of estimates. In spite of this, the \textsf{MLE} of parameters for size training data, varying from $20$ to $25$, are estimated and the future values over the span of the corresponding test data (where their sizes varying downward from $6-1$) are predicted. The \textsf{MAPE} of those predictions versus the size of training data are displayed in \cref{fig:sensitivity}. The largest \textsf{MAPE} is $3.18\%$, which is still very low, illustrating high quality forecasts, while demonstrating the robustness of our estimates to the choice of training data. 

\subsection{Step 2: 1 March--17 April 2020}\label{sec:PartII}

This step involves the first break point when the influence of lockdown restrictions appears in the COVID-19 population growth. The data from $1$ March--$3$ April $2020$ are used as the {\emph{training data}} to estimate the model parameters, using data from $4-17$ April $2020$ as the \emph{test data} to evaluate the accuracy of predictions. \cref{fig:cumsum_Aust} indicates changes to the dynamics of the growth rate after the first break point. Accordingly, we consider the following dynamics for the growth rate and the observation probability parameter for the underlying \texttt{PO-PBP}: 
\begin{align*}
	\lambda_t & = \begin{cases}
			\alpha_1^{t-t_0} \lambda & \mbox{for $t$ from $1-26$ March }2020, \\
			\alpha_2^{t-{bp_1}+1} \alpha_1^{bp_1-t_0} \lambda & \mbox{for $t$ from $27$ March--$17$ April }2020, 
		\end{cases}
\end{align*}
and
\begin{align*}
	p_t & = \begin{cases}
		\min\{\beta_1^{t-t_0} p, 1\} & \mbox{for $t$ from $1-26$ March }2020, \\
		\min\{ \beta_2^{t-{bp_1}+1} \beta_1^{bp_1-t_0} p , 1\} & \mbox{for $t$ from $27$ March--$17$ April }2020,
	\end{cases}
\end{align*}
where $\alpha_2 > 0$ and $\beta_2>0$ are new constant coefficients, and $bp_1$ is the date of the first break point (i.e., $27$ March $2020$). The two new parameters $\alpha_2$ and $\beta_2$ control the impact of the first break point on the growth rate and observation probability, respectively. \cref{tab:MLE_lambda_p-PartII} provides the \textsf{MLE} of the parameters. 
\begin{table}[h!]
	\begin{center}
		\begin{tabular}{ | l | c | c | c | c | c| c|}
			\hline
			\textbf{Parameters} & $\lambda$ & $\alpha_1$ & $\alpha_2$ & $p$ & $\beta_1$ & $\beta_2$  \\ \hline
			\textbf{\textsf{MLE}} & 0.235 & 1.000 & 0.814 & 0.67 & 1.00 & 1.06  \\ \hline
		\end{tabular}
		\caption{\textsf{MLE} of parameters}
		\label{tab:MLE_lambda_p-PartII}
	\end{center}
\end{table}

\begin{figure}[h!]
	\centering
	\includegraphics[width=12cm]{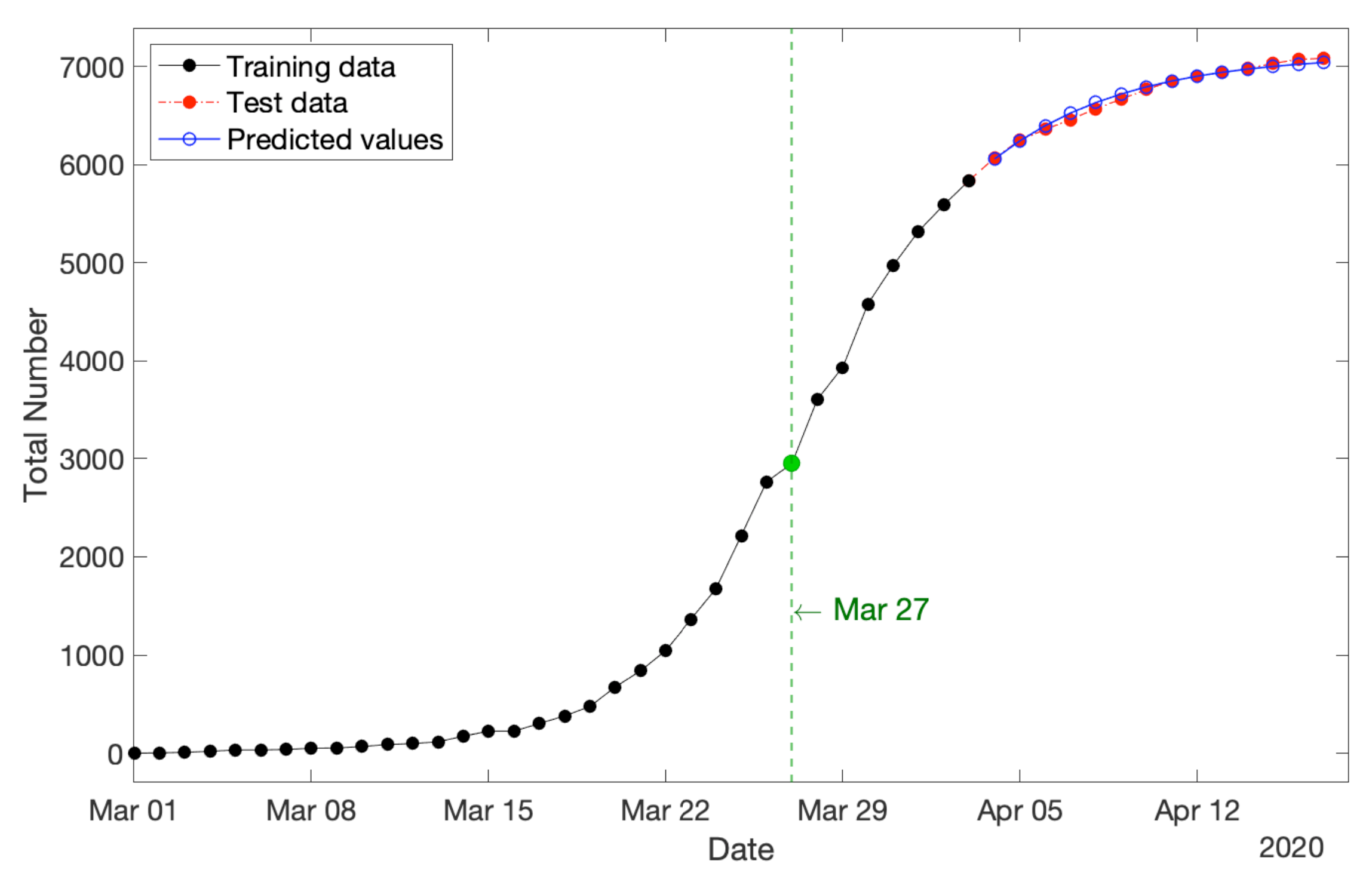}
	\caption{Cumulative new COVID-19 cases in Australia for three categories: (i) training data spanning $1$ March--$3$ April $2020$ (in black), test data spanning $4-17$ April $2020$ (in red), and (iii) predicted values over the period of test data (in blue) with $\mathsf{MAPE} = 0.43\%$. The first break point on $27$ March $2020$ is marked in green.}
	\label{fig:PartII}
\end{figure}

By applying \cref{pro:Ex_Yt}, the expected values of partial observations over $14$ days, that is the span of test data from $4-17$ April $2020$, are predicted. The results are shown in \cref{fig:PartII}, where the training data, test data, and predictions are displayed by the black solid plot, red dot-dash plot, and blue solid plot, respectively, and the first break point is marked in green. Clearly, the predicted values are remarkably fitted to the actual test data with a significantly small $\mathsf{MAPE} = 0.43\%$, which is notably less than one percent error. 

\begin{remark}
	The \textsf{MLE}s provided in \cref{tab:MLE_lambda_p-PartII} imply that the observation probability $p_t$ starts increasing after the first break point with the estimated boosting factor of $\mathsf{MLE}(\beta_2) = 1.06$ such that after one week, it reaches the upper bound of $1$. However, by considering the identifiability of these point estimations as well as \cref{rem:rule_of_thumb}, it is observed that the locus allows for $\mathsf{MLE}(p_t)$ to range within $[0.95,1.00]$ over the test period of $4-17$ April $2020$. Hence, the model estimates that the percentage of observed COVID-19 cases from $2-17$ April $2020$ lies within the range $[95\%,100\%]$. Furthermore, analogous to \cref{fig:sensitivity}, the robustness of estimates is confirmed. 
\end{remark}

\subsection{Step 3: 1 March--22 May 2020}\label{sec:PartIII}

The last step is for the whole span of the data, involving both structural break points. The data from $1$ March--$24$ April $2020$ are used as the {\emph{training data}} to estimate the parameters of the model, and the data from $25$ April--$22$ May $2020$ are used as the \emph{test data} to evaluate the quality of predictions. Motivated from \cref{fig:cumsum_Aust} along with Steps 1--2, we define the following dynamics for the growth rate and the observation probability parameter for the underlying \texttt{PO-PBP}: 
\begin{align*}
	\lambda_t & = \begin{cases}
		\alpha_1^{t-t_0} \lambda & \mbox{for $t$ from $1-26$ March }2020, \\
		\alpha_2^{t-{bp_1}+1} \alpha_1^{bp_1-t_0} \lambda & \mbox{for $t$ from $27$ March--$17$ April }2020, \\
		\alpha_3^{t-{bp_2}+1} \alpha_2^{bp_2-bp_1} \alpha_1^{bp_1-t_0} \lambda & \mbox{for $t$ from $18$ April--$22$ May }2020, 
	\end{cases}
\end{align*}
and
\begin{align*}
	p_t & = \begin{cases}
		\min\{\beta_1^{t-t_0} p, 1\} & \mbox{for $t$ from $1-26$ March }2020, \\
		\min\{ \beta_2^{t-{bp_1}+1} \beta_1^{bp_1-t_0} p , 1\} & \mbox{for $t$ from $27$ March--$17$ April }2020, \\
		\min\{ \beta_3^{t-{bp_1}+1} \beta_2^{bp_2-bp_1} \beta_1^{bp_1-t_0} p , 1\} & \mbox{for $t$ from $18$ April--$22$ May }2020,
	\end{cases}
\end{align*}
where $\gamma > 0$ is the new inflation parameter on the growth rate after the second break point, and $bp_2$ is the date of the second break point (i.e., $18$ April $2020$). \cref{tab:MLE_lambda_p-PartIII} provides the \textsf{MLE} of the parameters.
\begin{table}[h!]
	\begin{center}
		\begin{tabular}{ | l | c | c | c | c | c| c| c| c |}
			\hline
			\textbf{Parameters} & $\lambda$ & $\alpha_1$ & $\alpha_2$ & $\alpha_3$ & $p$ & $\beta_1$ & $\beta_2$ & $\beta_3$ \\ \hline
			\textbf{\textsf{MLE}} & 0.235 & 1.000 & 0.820 & 0.979 & 0.67 & 1.00 & 1.06 & 1.00 \\ \hline
		\end{tabular}
		\caption{\textsf{MLE} of parameters}
		\label{tab:MLE_lambda_p-PartIII}
	\end{center}
\end{table}

\begin{figure}[h!]
	\centering
	\includegraphics[width=12cm]{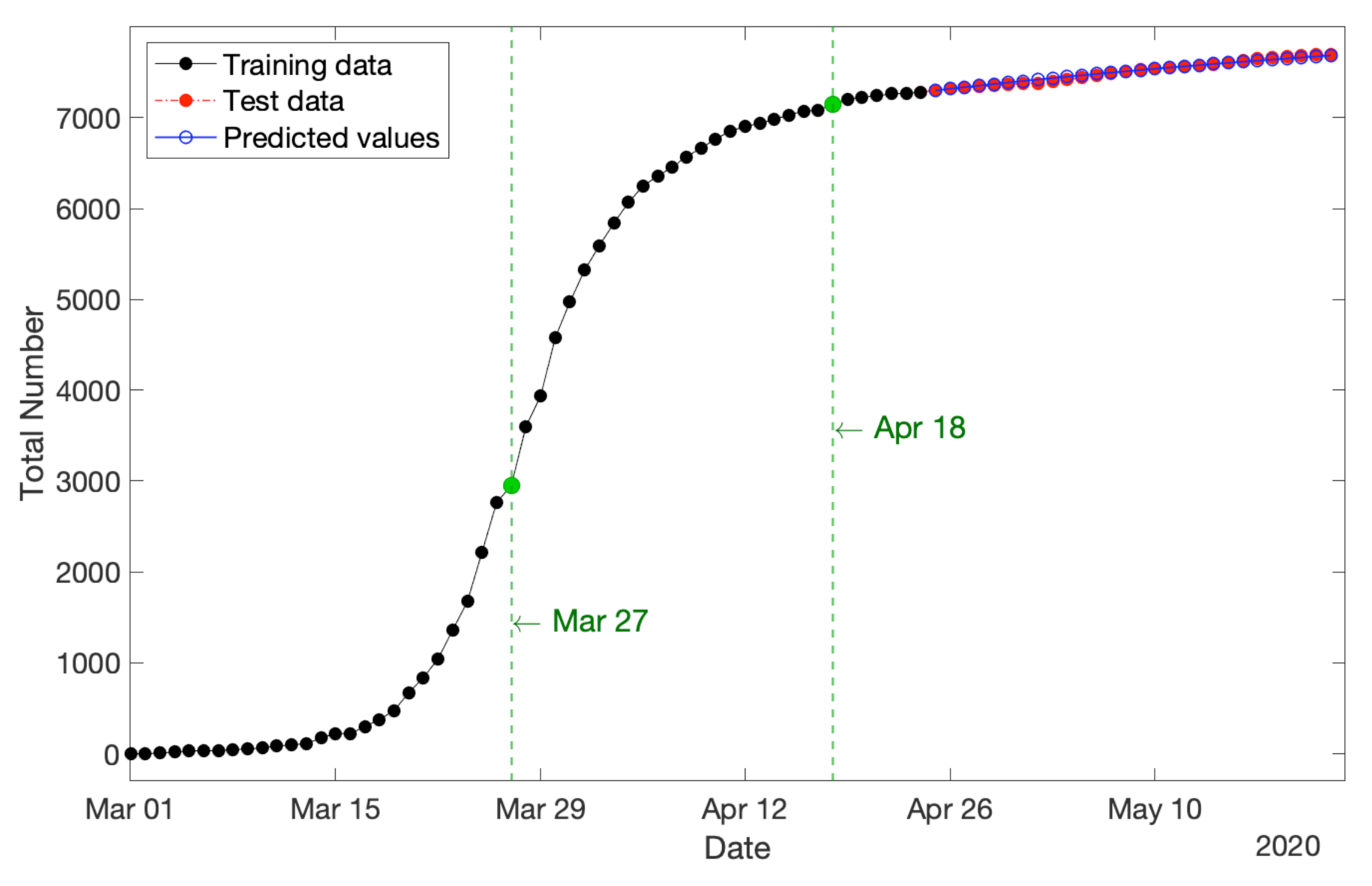}
	\caption{Cumulative new COVID-19 cases in Australia for three categories: (i) training data spanning $1$ March--$24$ April $2020$ (in black), test data spanning $25$ April--$22$ May $2020$ (in red), and (iii) predicted values over the period of test data (in blue) with $\mathsf{MAPE} = 0.20\%$. The two break points are marked in green.}
	\label{fig:PartIII}
\end{figure}

By applying \cref{pro:Ex_Yt}, the expected values of partial observations over the span of test data (i.e., $28$ days from $25$ April--$22$ May $2020$, inclusive) are predicted. Results are shown in \cref{fig:PartIII}, where the training data, test data, and predictions are displayed by the black solid plot, red dot-dash plot, and blue solid plot, respectively, and the two break points are marked in green. It is clearly evident that the predicted values are exceptionally closely fitted to the actual test data with $\mathsf{MAPE} = 0.20\%$, which is notably much less than one percent. 

\begin{remark}
	The \textsf{MLE}s provided in \cref{tab:MLE_lambda_p-PartIII} show that the declining parameter on the growth rate is inflated from $\mathsf{MLE}(\alpha_2) = 0.820$ to $\mathsf{MLE}(\alpha_3) = 0.979$. Although it is still less than one (indicating that the population is stable and not exploding), but such an increase as a consequence of people's interactions during the Easter break as well as releasing a few restrictions on $2$ May $2020$ should be taken into account by policy makers for easing out the COVID-19 restrictions. Furthermore, by considering the identifiability of estimations as well as \cref{rem:rule_of_thumb}, it is observed that the locus allows for $\mathsf{MLE}(\alpha_3)$ to range within $[0.940,1.020]$. Hence, there is a chance that the parameter $\mathsf{MLE}(\alpha_3)$ could be greater than one, implying that the population starts growing again. If this takes place, the population size will quickly resume exponential growth (cf. \cref{fig:PartI_2}). Furthermore, analogous to \cref{fig:sensitivity}, the robustness of estimates is confirmed. 
\end{remark}

 \begin{remark}\label{rem:algorithm_for_structural-break-points}
 	One can easily see that the \texttt{MLE}s of parameters provided in \cref{tab:MLE_lambda_p-PartI,tab:MLE_lambda_p-PartII,tab:MLE_lambda_p-PartIII}, alter after each per-identified structural break point. These results confirm our choice of break points, and also motivate us to suggest an algorithmic way to detect the break points. In that case, the whole time frame can be partitioned into $\tau$ mutually exclusive time intervals $(t_{k-1}, t_k)$ for $k=1,\ldots,\tau$, and the following dynamics for the growth rate parameter $\lambda_t$ and the observation probability $p_t$ can be constructed:
 	\begin{align*}
 	\lambda_t & = \begin{cases}
 	\alpha_1^{t-t_0} \lambda & \mbox{for } t\in [t_0,t_1), \\
 	\Pi_{j=1}^{k-1} \alpha_j^{t_j-t_{j-1}} \alpha_k^{t-t_{k-1} + 1} \lambda & \mbox{for } t \in [t_{k-1}, t_{k}),\ j=2,\dots,\tau,
 	\end{cases}
 	\end{align*}
 	and
 	\begin{align*}
 	p_t & = \begin{cases}
 	\min\{\beta_1^{t-t_0} p, 1\} & \mbox{for } t\in [t_0,t_1), \\
 	\min\{\Pi_{j=1}^{k-1} \beta_j^{t_j-t_{j-1}} \beta_k^{t-t_{k-1} + 1} p, 1\} & \mbox{for } t \in [t_{k-1}, t_{k}),\ j=2,\dots,\tau.
 	\end{cases}
 	\end{align*}
 	After estimating all parameters, those consecutive intervals showing distinct \textsf{MLE}s for $\alpha_i$ and $\beta_i$ could be an indication of a ``structural break point''. If one wants to trim estimating the location of each break point, one should merge those consecutive intervals first, then partition them into some more sub-intervals with new parameters, and re-estimate all parameters together. Such trimming procedure can be repeated until there is no change in the break points. 
 \end{remark}

%% file: Strengths_and_Limitations.tex
\section{{PO-PBP}: Strengths and Limitations} \label{sec:Strengths_and_Limitations}

The main strength of the \texttt{PO-PBP} model presented in this paper is the high accuracy of predictions within a timescale in the order of $4$ weeks. It is timescales of this order that governments are considering to adjust restrictions and modify adjustments to restrictions\footnote{\url{https://www.pm.gov.au/media/update-coronavirus-measures-08may20}}. Such high accuracy on these timescales makes the model not only applicable but also highly appealing to base policy decisions on.

A natural question to ask is whether comparable accuracy can be obtained in this work by directly fitting a pure birth process to the data-segments that lie between break points. We have made this calculation and conducted the comparison with our \texttt{PO-PBP} model and found that although the \texttt{PBP} performs reasonably well, it is not as accurate as our \texttt{PO-PBP}. For instance, the \textsf{MAPE} for predictions before the lockdown restrictions obtained by the former model is in the order of $6.03\%$, which is not as good as the \textsf{MAPE} for our \texttt{PO-PBP}, which is only $1.53\%$.

Would the accuracy of predictions of this model extend into the long-range future?  This model, as with any models, should be used within its range of validity, with care to take into account the assumptions built into it. Due to some of the assumptions in the model that are described below, long-range predictions will lose accuracy unless they are corrected for by updated real-life data.  

The model in this paper is based on a \texttt{PBP}, which is a special case of a more general Continuous-time Markov population process that includes both births and deaths. In this case, a ``birth'' is a new infection, and a ``death'' represents removal from the infected population by means of either recovery from illness or death. The real-life process of COVID-19 infection includes both infection and potential recovery or death. Thus in that sense a richer model that includes both birth and death would be appropriate.  However, there is a trade-off. The theory for the richer model exists and is sound, but is computationally much less tractable.  

The trade-off between representational accuracy and computability is a common one in mathematical modeling, and well understood in the literature (e.g., see \cite{Tibshirani}). In the case of birth and death processes, if births happen much faster than deaths, it is legitimate to disregard deaths within an appropriate timescale. In the case of COVID-19, infection can happen very quickly -- in a matter of hours, whereas recovery or potentially death takes much longer: weeks potentially stretching into a month or more. Thus on timescales of a few weeks, much insight can be gained from a \texttt{PBP}.

Furthermore, the trade-off between the tractability of a \texttt{PBP} versus the longer range accuracy of one utilizing both births and deaths can be further tipped in the balance towards pure birth modeling by a method which to some extent accounts for both processes. Consider a growing birth and death process with the birth and death rate of $\lambda$ and $\mu$, respectively, such that $\lambda > \mu$. Then to some extent, both may be accounted for in a \texttt{PBP} in which birth rate is modeled by the difference $\lambda - \mu$. Our \texttt{PO-PBP} model employs this strategy: the $\lambda$ in our model is really the difference between an underlying birth and death rate.  This extends slightly the timescales within which the model is useful.

The considerations above are relevant to the prediction in \cref{fig:PartI_2}, in which the blue line indicates that if growth rates had continued in the same pattern as pre-lockdown, all Australians would have been infected by $24$ April $2020$. In reality that pattern would have been somewhat mitigated by factors not currently present in our \texttt{PO-PBP}. One such factor is recovery/death rates discussed above. Another is the finite size of the Australian population and our island status.

Due to Australia being an island along with recently closing its borders, any circulating virus has only a finite collection of approximately $25$ million people to potentially infect. The more people infected, the greater the chance that when an infected person comes into contact with another person, that other person is also already infected, so that no new infection can occur. This saturation effect is not built into our \texttt{PO-PBP}, which means that the blue line in \cref{fig:PartI_2} would shoot straight off the page had we not truncated it. In alternative models that account for the limiting effect of finite population size, that blue line would have curved down slightly as the infected proportion of the population became comparable with the overall population size.  In other words, the domain of validity of the present \texttt{PO-PBP} is limited to the situation in which the overall proportion of infected individuals is still relatively low, as is presently the case, as of May $2020$.

At this point, it is insightful to compare with another form of disease modeling, the use of \texttt{SIR} models utilizing differential equations. The acronym ``SIR'' stands for three categories, respectively: Susceptible, Infected and Removed. A \emph{susceptible} person has not yet caught the infection, but potentially could; an \emph{infected} person actively has the infection, and is assumed capable of passing it on; and a \emph{removed} person is removed from the population of susceptible or infected people by either death or immunity. In basic \texttt{SIR} modeling, a flow diagram is drawn between the three categories, and the flows between these are modeled by differential equations in which parameters are introduced to represent the chances of transitioning between the states. In this way the inter-relationships between the categories are captured. Solutions to the equations predict how the size of the three categories will develop over time. These solutions may be obtained either exactly or numerically, and sometimes depend critically on accurate values of the parameters in the \texttt{SIR} model.

An advantage of the \texttt{SIR} model over our \texttt{PO-PBP} is that the \texttt{SIR} accounts for the finite population size, and removes persons who have become immune or died from the susceptible/infected population. The disadvantage, however, is that the \texttt{SIR} model does not explicitly account for the ``partially observed'' nature of COVID-19 cases. Policy makers and Scientists only have access to reported data, yet there may well be infections in the population which are driving the growth of the pandemic, but which are not directly diagnosed and hence observed.

The most remarkable advantage of the \texttt{PO-PBP} is that it provides means of estimating and incorporating the proportion of hidden cases. More precisely, our model employs a new ``observation probability'' parameter to the underlying \texttt{PBP} to construct the new \texttt{PO-PBP}. Then, by maximizing the complicated likelihood function of the \texttt{PO-PBP}, all parameters including the observation probability at each time $t$ are estimated. Due to the invariant property of \textsf{MLE}s, one minus the \textsf{MLE} of that observation probability will estimate the proportion of hidden cases in the population. 

Any model, however, is only as valid as its assumptions. An assumption of the \texttt{PO-PBP} is that sampling is ``uniform and random'', implying that the model assumes any infected person as likely to be tested and identified as any other person. The assumption of randomness is almost never $100\%$ satisfied for any realistic scenario -- some biases will always be present. What matters is how impactful these are. It is worth considering the impact of model assumptions in this modeling.  

One feature of relevance is the availability of test kits. If shortage of test kits were to severely curtail sampling, this would undermine the validity of the model. In Australia, initial testing was largely limited to people considered ``high risk'', and include those who recently returned from overseas, had contact with a confirmed COVID-19 case, or in hospital with severe symptoms matching the disease, while other population members were considered ``low risk''. If perchance significant infection had established in the ``low risk'' part of the population, and if a low death rate had allowed this hidden population to remain undetected, the proportion of undetected cases reported by our model would be an underestimate. However, the predicted pattern of confirmed infections would remain valid.

Recently, the Australian government has substantially expanded testing opportunities and let testing for COVID-19 be available to every Australian with mild respiratory symptoms including a cough and sore throat\footnote{\url{https://www.pm.gov.au/media/update-coronavirus-measures-24april20}}. This makes the ``random sampling'' assumption of the \texttt{PO-PPB} considerably more robust. We should soon be able to determine whether there has been a significant reservoir of undetected COVID-19 cases in Australia. 

It should be noted that we are implementing a continuous-time model, whilst observations are reported just once daily. Use of a continuous-time model is still valid, however, since it is general enough to take account of the discrete data structure. This continuous-time model is utilized due to the power of the theory that underlines the model and the relevance of that theory to this modeling situation.  

We conclude this section by stating that the \texttt{PO-PBP} models only the impact of COVID-19 with respect to the numbers of infections -- it does not model other impacts on society (negative or positive) of policy control measures, as it is recognized that restrictions on gatherings affect people's lives in different ways. A positive example is reduced air pollution due to reduced travel \cite{Dutheil2020}, while a negative example is increased risk of harms like domestic violence \cite{BJ20}. As yet there is no single model which incorporates all of these factors.

%% file: Discussion_and_Conclusion.tex
\section{Discussion and Conclusion: Policy Implications} \label{sec:Discussion_and_Conclusion}

In this paper, we apply a class of continuous-time Markov population stochastic processes, namely the \emph{partially-observable pure birth process}, to model and analyze the dynamics of the COVID-19 population in Australia. Specifically, we use the theoretical properties of this stochastic process to construct the likelihood function of cumulative confirmed cases to find the maximum likelihood estimates of its parameters. These estimations are used to predict future values of the population along with the number of unobserved hidden COVID-19 cases. 
 
The Markovian stochastic process model that we develop is based on a partially observable pure birth process, and its predictions fit the actual observations at the Australian national level surprisingly well. Aside from its simplicity and high accuracy, there are several other advantages of this model from a policy perspective. 

The stochastic process in our model revolves around only two parameters, both of which have clear and communicable practical interpretations: the former represents the speed of the spread, while the latter provides a measure of detection likelihood. We postulate that in the absence of any policy interventions, both parameters follow an evolution pattern that resembles a geometric decay. A shift in policy, however, may ratchet up or down the decay rate, thereby inducing a new infection trajectory.  

As demonstrated, the model captures the complex dynamics of the detected/infected ratio, which is a critical component in the design of any containment policy. Furthermore, policymakers gain access to a coherent and insightful representation of the situation to informatively contemplate the consequences of action/inaction over the span of a few weeks (i.e., how the epidemic unfolds in the absence of any interventions). 

Because of its efficacy, this model equips policymakers with a powerful tool to conduct scenario-based analyses, and enables predictions with a high degree of precision, of how a particular decision drifts the evolution trajectory of the disease, and enables such a prediction only a week after it is enacted. This supports early identification and reinforcement of effective policies, and timely scale back or discontinuation of others. 

The model empowers decision-makers to evaluate and compare the implications of the two fundamental hallmarks of the model: lowering the infection rate versus increasing the detection likelihood. Depending on which one of these two avenues should be pursued, the government resources should be directed accordingly, and the corresponding message should be conveyed to the public. 

If extensive community screening were undertaken particularly for asymptomatic citizens, then this model would be expected to give very accurate predictive power throughout the full range of possible policy implementation with regard to social distancing restrictions. The model further lends itself to crafting hybrid policies that utilize a combination of these two approaches and the delicate division of available resources between them. 

Availability of test kits is a practical consideration in interpreting this model. The current model assumes an adequate supply of test-kits so that sampling is not restricted by a shortage. If this assumption was not fully met during the early days of the pandemic in Australia, it would mean we would have potentially underestimated the ``hidden fraction'' of undetected COVID-19 cases. Statements by the government\footnote{\url{https://www.linkedin.com/in/scottmorrisonmp/}} indicate that there is no current shortage of test kits in Australia.  It would be interesting but quite difficult future work to try to explicitly incorporate test-kit shortage in the model. Such future work would have particular relevance in other countries where test-kit shortages are a pressing issue.   

Additionally, our epidemic model treats the entire nation as a single pool of homogeneous agents with equal exposure to risks and similar contact behaviors. While this assumption may sound fairly restrictive, we believe that it has not impacted the quality of our findings in a profound way. Nevertheless, accounting for inherent heterogeneities and local characteristics of smaller regions/communities would further enhance the richness of the model and strengthen its outcomes. This modification particularly lends itself to countries such as U.S. where there is substantial variation in the extent and timing of the epidemic across states.

This model is likely to be applicable to many other countries and circumstances beyond Australia, since there are not a large number of location-specific assumptions built in, the main one being that testing involves a reasonably uniform sampling of the population. This strength partly derives from the analytical nature of the model, rather than being one which is simulation-based and within that highly customized to local factors. Ideally, different models are used in conjunction, and this model could profitably be used in conjunction with other types of models, in understanding the spread of COVID-19 in the future both in Australia and elsewhere.  

A benefit of this model is that the $4$ weeks prediction horizon allows officials to fine-tune their short-term actions and contingency planning in light of reasonable confidence in the immediately expected upcoming pattern in the number of cases. 

This model suggests several possibilities for further research, that would enhance its applicability across a broader range of circumstances. For example, in the Australian data, the structural break-points were identified visually. There is potential to develop a purely computational method of doing these identifications, according to \cref{rem:algorithm_for_structural-break-points}. This would enhance applicability to more complex and long-term data sets, as may be expected in the future across the world.

All in all, the \texttt{PO-PBP} is a useful model for understanding and predicting the trajectory of COVID-19 in Australia under various policy choices. On a short timescale, relevant to government actions, predictions have been shown to be very accurate. The model also appears to be sensitive to subtle shifts in population behavior, allowing it to be useful in considering the impact of social events such as the Easter break in April $2020$.

%% file: Acknowledgment.tex
\paragraph{Acknowledgment}

The authors are very grateful to Dr Judy-anne Osborn from the University of Newcastle in Australia for extensive useful conversations about modeling assumptions, implications and presentation thereof, during preparation of this paper. The authors also thank Prof. Nigel Bean from the University of Adelaide in Australia for suggesting some improvements.

%% file: Appendix.tex
\appendix
\section{Appendix: Background} \label{sec:Appendix}

In this appendix, we present a brief overview of the underlying stochastic process used for modeling in this paper. An extensive discussion can be found in \cite{BEER_POCMP,BER_FI}.

\subsection{Partially-observable Continuous-time Markov Population Process}\label{sec:POCTMPM}

Suppose $\{X_t,\,t\geq 0\}$ is a continuous-time Markov population process with the unknown parameter vector $\bm{\theta}_t$. The vector $\bm{\theta}_t$ parameterizes the q-matrix (generator) $Q(\bm{\theta}_t)$ of the model. We restrict our attention to \texttt{CTMPP}s where the range of the random variable $X_t$ includes non-negative integers, and the initial value of this process, $x_0$, is known. Moreover, we suppose that the process is time-homogeneous, that is the conditional probability $\mathsf{P}_{(X_{t_2}|X_{t_1})}(x_{t_2}|x_{t_1})$ for any values of $t_2>t_1\geq 0$ depends only on $x_{t_1}$, $x_{t_2}$ and $t_2-t_1$.

In order to estimate the unknown parameter vector $\bm{\theta}_t$, we take $n$ observations of $\{X_t,\,t\geq 0\}$ at times $0< t_1\leq t_2\leq \cdots\leq t_n$. Suppose that at each observation time $t_i$, we do not observe $X_{t_i}$ directly, but rather only a random sample. This may be due to practical restrictions such as time or budget constraints which limit the ability to survey comprehensively, or might be because of an implicit component of the data collection process. A common model for the sampling is binomial, where the state of the system, or each component of the system, is observed with a probability $p_t$ at observation time $t$. \cref{def:PO-CTMPP} provides a formal definition of a \emph{partially-observable continuous-time Markov population process}. 

\begin{definition}[\cite{BEER_POCMP}]\label{def:PO-CTMPP}
	Consider the \texttt{CTMPP} $\{X_t,\,t\geq 0\}$ with the parameter vector $\bm{\theta}_t$. Suppose the random variables $Y_t$ are defined such that the conditional random variable $(Y_t|X_t=x_t)$ follows the \textsf{Binomial}$(x_t,p_t)$ distribution, that is
	\begin{align*}
	\mathsf{P}_{(Y_t|X_t)}(y_t|x_t) & = \binom{x_t}{y_t} p_t^{y_t} (1-p_t)^{x_t-y_t}\ \ \ \mbox{for}\ y_t=0,1,\ldots,x_t\,.
	\end{align*}
	Then the stochastic process $\{Y_t,\,t\geq 0\}$ is called a \texttt{PO-CTMPP} with the parameter vector $(\bm{\theta}_t,p_t)$.
\end{definition}

\begin{remark}
	It is readily seen that a \texttt{PO-CTMP} model with parameter vector $(\bm{\theta}_t,1)$ reduces to a \texttt{CTMP} model with parameter vector $\bm{\theta}_t$.
\end{remark}

In order to find the \textsf{MLE} of the unknown parameter vector $(\bm{\theta}_t,p_t)$, we first need to construct the likelihood function of partial observations, that is, 
\begin{align*}
\medskip \mathsf{L}_{\bm{Y}_n}(\bm{y}_n;\bm{\theta}_t,\bm{p}_n) & = \Pr(\bm{Y}_n=\bm{y}_n),
\end{align*}
where the random vector $\bm{Y}_n:=(Y_0, Y_{t_1}, Y_{t_2}, \cdots, Y_{t_n})$, the realization vector $\bm{y}_n:=(x_0, y_{t_1}, y_{t_2}, \cdots, y_{t_n})$, the probability vector $\bm{p}_n:=(1, p_{t_1}, p_{t_2}, \cdots, p_{t_n})$, and $\Pr(Y_{0} = x_0) = 1$.

Bean et al.\cite{BEER_POCMP} utilized the Conditional Bayes' Theorem \cite{Eli} and derived the following analytical results: 

\begin{theorem}[\cite{BEER_POCMP}] \label{thm:PO-CTMP}
	Consider a \texttt{PO-CTMP} process with the parameter vector $(\bm{\theta}_t,p_t)$. 
	\begin{enumerate}[label = (\roman*)]
		\item \medskip The conditional p.m.f. of the true value of the underlying process given the partial observations is 
		\begin{align*}
		\mathsf{P}_{(X_{t_n}|\bm{Y}_n)}(x_{t_n}|\bm{y}_n) & = \frac{\varrho_n^{x_{t_n}}}{\displaystyle{\sum_{\ell=y_{t_n}}^{\infty}\varrho_n^{\ell}}}\ \ \mbox{for}\ x_{t_n}=y_{t_n},y_{t_n}+1,\ldots,
		\end{align*} 
		where,
		\begin{align*}
		\varrho_n^{\ell} & := e y_{t_n}!\binom{\ell}{y_{t_n}} p_{t_n}^{y_{t_n}} (1-p_{t_n})^{\ell-y_{t_n}} \sum_{j=y_{t_{n-1}}}^{\infty} \mathsf{P}_{(X_{t_n}|X_{t_{n-1}})}(\ell|j) \varrho_{n-1}^{j},
		\end{align*}
		for $\ell=y_{t_n},y_{t_n}+1,\ldots$, $n=1,2,\ldots$, and the initial conditions $\varrho_{0}^{x_0}=1$ and $\varrho_{0}^{\ell}=0$ for $\ell \neq x_0$.
		
		\item \medskip The conditional p.m.f. $\mathsf{P}_{(Y_{t_{n+1}}|\bm{Y}_n)}(y_{t_{n+1}}|\bm{y}_n)$ for $y_{t_{n+1}}=0,1,2,\ldots$ equals to
		\begin{align*}
		& \frac{\displaystyle{\sum_{x_{t_{n+1}}=y_{t_{n+1}}}^{\infty} \sum_{x_{t_{n}}=y_{t_n}}^{\infty} \binom{x_{t_{n+1}}}{y_{t_{n+1}}} p_{t_{n+1}}^{y_{t_{n+1}}} (1-p_{t_{n+1}})^{x_{t_{n+1}}-y_{t_{n+1}}} \mathsf{P}_{(X_{t_n}|X_{t_{n-1}})}(x_{t_{n}}|x_{t_{n-1}}) \varrho_n^{x_{t_n}}}} {\displaystyle{\sum_{\ell=y_{t_n}}^{\infty}\varrho_n^{\ell}}},
		\end{align*}
		for $n=1,2,\ldots$.
	\end{enumerate}
\end{theorem}

\subsection{Partially-observable Pure Birth Process}\label{sec:POPBP}

A popular model in the class of \texttt{CTMPP} is the stochastic \emph{pure birth process} (\texttt{PBP}). Let $\{X_t,\,t\geq 0\}$ be a time-homogeneous \texttt{PBP}, with the parameter $\lambda_t$ (known as the birth/growth rate) at time $t$, and known initial population size of $x_0$. If $X_t=x_t$, then the \emph{transition rate} equals to $\lambda_t x_t$. It can be shown \cite{Ren91} that if the birth rate over a given time interval $[t_1,t_2]$ does not vary and equals to $\lambda_{t_1}$, then the transition probability at times $0\leq t_1 \leq t_2$ is given by
\begin{align*}
\medskip \mathsf{P}_{(X_{t_2}|X_{t_1})}(x_{t_2}|x_{t_1}) & = \binom{x_{t_2}-1}{x_{t_1}-1} e^{-\lambda_{t_1}(t_2-t_1)x_{t_1}} (1-e^{-\lambda_{t_1}(t_2-t_1)})^{x_{t_2}-x_{t_1}}\ \ \mbox{for}\ x_{t_2} = x_{t_1},x_{t_1}+1,\ldots.
\end{align*}

Let the stochastic process $\{Y_t,\,t\geq 0\}$ be the corresponding \emph{partially-observable pure birth process}(\texttt{PO-PBP}), with the parameter vector $(\lambda_t,p_t)$. Bean et al. \cite{BEER_POCMP} simplified Theorem \cref{thm:PO-CTMP} for a \texttt{PO-PBP}, as provided in \cref{cor:PO-PBP}.

\begin{corollary}[\cite{BEER_POCMP}]\label{cor:PO-PBP}
	Consider a \texttt{PO-PBP} $\{Y_t,\ t\geq 0\}$ with the parameter vector $(\lambda_t,p_t)$, and the underlying \texttt{PBP} $\{X_t,\ t\geq 0\}$ with the known initial population size of $x_0$. 
	
	\begin{enumerate} [label = (\roman*)]
		\item \medskip The quantity $\varrho_n^{\ell}$ for $\ell=y_{t_n},y_{t_n}+1,\ldots$, and $n=1,2,\ldots$, is given by
		\begin{align*}
		& e y_n!\binom{\ell}{y_n} p_{t_n}^{y_n} (1-p_{t_n})^{\ell-y_n} \sum_{j=\overline{x}_t{_{n-1}}}^{\ell}   \binom{\ell-1}{j-1} e^{-\lambda_{t_{n-1}}(t_n-t_{n-1}) j} (1 - e^{-\lambda_{t_{n-1}}(t_n-t_{n-1})})^{\ell-j} \varrho_{n-1}^{j},
		\end{align*}
		where $\overline{x}_{t_{n}}:=\max\{x_0,y_{t_1},\cdots,y_{t_{n}}\}$. The initial conditions are as provided in \cref{thm:PO-CTMP}. 
		
		\item \medskip The conditional p.m.f. of the next partial observation, given all past $n$ partial observations equals to
		\begin{align*}
		\mathsf{P}_{(Y_{t_{n+1}}|\bm{Y}_n)}(y_{t_{n+1}}|\bm{y}_n) & = \frac{1}{\displaystyle{\sum_{\ell=\overline{x}_{t_{n}}}^{\infty}\varrho_n^{\ell}}} \left(\displaystyle{\sum_{x_{t_{n+1}}=\overline{x}_{t_{{n+1}}}}^{\infty} \sum_{x_{t_{n}}=\overline{x}_{t_{n}}}^{x_{t_{n+1}}} \binom{x_{t_{n+1}}}{y_{t_{n+1}}} p_{t_{n+1}}^{y_{t_{n+1}}} (1-p_{t_{n+1}})^{x_{t_{n+1}}-y_{t_{n+1}}}}\right. \\
		& \left.\times \displaystyle{\binom{x_{t_{n+1}}-1}{x_{t_n}-1} e^{-\lambda_{t_n}(t_{n+1}-t_n)x_{t_n}} (1-e^{-\lambda_{t_n}(t_{n+1}-t_n)})^{x_{t_{n+1}}-x_{t_n}} \varrho_n^{x_{t_n}}}\right),
		\end{align*}
		for $y_{t_{n+1}}=0,1,2,\ldots$, and $n=1,2,\ldots$.
	\end{enumerate}	
\end{corollary}

An important question that may arise here is the dependency structure of the stochastic process $\{Y_t,\ t\geq 0\}$ which is addressed in $t\in(0,\infty)$. \cref{thm:PO-nonMarkovian}.

\begin{theorem}[\cite{BEER_POCMP}]\label{thm:PO-nonMarkovian}
	The \texttt{PO-CTMP} process is not Markovian of any order. That is, for any fixed value of $k=1,2,\ldots$, there exist $0\leq t_1\leq t_2\leq \cdots\leq t_n$, $y_1, y_2, \cdots, y_n$, and $n>k$, such that,
	\begin{align*}
	& \Pr(\left.Y_{t_n}=y_{t_n}\right|Y_{t_1}=y_{t_1},\cdots,Y_{t_{n-1}}=y_{t_{n-1}}) \neq \Pr(Y_{t_n}=y_{t_n}|Y_{t_{n-k}}=y_{t_{n-k}},\cdots,Y_{t_{n-1}}=y_{t_{n-1}}).
	\end{align*} 
\end{theorem} 

\paragraph*{Likelihood function. }Although, \cref{thm:PO-nonMarkovian} makes finding the likelihood function of a \texttt{PO-PBP} more challenging and complicated, one can use the chain rule along with \cref{cor:PO-PBP} to construct the likelihood function:
\begin{align}\label{eq:LF-chain_rule}
\medskip \mathsf{L}_{\bm{Y}_n}(\bm{y}_n; \bm{\lambda}_n,\bm{p}_n) & = \prod_{k=1}^{n} \mathsf{P}_{(Y_{t_{k}}|\bm{Y}_{k-1})}(y_{t_{k}}|\bm{y}_{k-1}),
\end{align} 
where $\bm{\lambda_n} := (\lambda_0, \lambda_{t_1},\cdots, \lambda_{t_n})$.
Now, by having the likelihood function at hand, one can find the \textsf{MLE} of unknown parameters for a \texttt{PO-PBP}. However, there are some infinite sums involved with the likelihood function which should be handled carefully in numerical computations. One approach to deal with those infinite sums is to truncate them by exploiting Chebyshev's inequality. More precisely, Chebyshev's inequality prescribes to truncate the infinite sum over the realizations of the conditional random variable $(X_{t_n} \mid \bm{Y}_{n}=\bm{y}_{n})$ at
\begin{align}\label{eq:truncation}
& \Ex\left[X_{t_n} \mid \bm{Y}_{n}=\bm{y}_{n}\right] + 20 \sqrt{\mathsf{Var}(X_{t_n} \mid \bm{Y}_{n}=\bm{y}_{n})}\,,
\end{align}
to guarantee that at least $99.75\%$ of the corresponding probability distribution is covered. Been et al. \cite{BER_FI} derived those expected values involved in the truncation point \cref{eq:truncation} analytically. 

\begin{proposition}[\cite{BER_FI}]\label{pro:Ex_Var}
	Consider a \texttt{PO-PBP} $\{Y_t,\ t\geq 0\}$ with the parameter vector $(\lambda_t,p_t)$, and the underlying \texttt{PBP} $\{X_t,\ t\geq 0\}$. We have,
	\begin{align*}
	\medskip \Ex\left[X_{t_n} \mid \bm{Y}_{n}=\bm{y}_{n}\right] & = \frac{\overline{x}_{t_n} + (1-p_{t_n})(1-e^{-\lambda_{t_n} {t_n}})}{p_{t_n} + (1-p_{t_n})e^{-\lambda_{t_n} {t_n}}}, \\
	\medskip \mathsf{Var}(X_{t_n} \mid \bm{Y}_{n}=\bm{y}_{n}) & = \frac{(\overline{x}_{t_n}+1) (1-p_{t_n}) (1-e^{-\lambda_{t_n} {t_n}})}{(p_{t_n} + (1-p_{t_n})e^{-\lambda_{t_n} {t_n}})^2},
	\end{align*}
	where $\overline{x}_{t_n}$ is as defined in \cref{cor:PO-PBP}.
\end{proposition}

\paragraph*{Prediction. }In order to predict the future values of the process given the past partial observations, we use the \textsf{MLE} of the conditional expected value $\Ex\left[Y_{t_{n+1}} \mid \bm{Y}_{n}=\bm{y}_{n}\right]$. Due to the invariant property of \textsf{MLE}s, we only need to find the \textsf{MLE} of the unknown parameters $\lambda_t$ and $p_t$ and replace them in the equation provided in \cref{pro:Ex_Yt}. 

\begin{proposition}[\cite{BER_FI}]\label{pro:Ex_Yt}
	Consider a \texttt{PO-PBP} $\{Y_t,\ t\geq 0\}$ with the parameter vector $(\lambda_t,p_t)$, and the underlying \texttt{PBP} $\{X_t,\ t\geq 0\}$. We have,
	\begin{align*}
	\medskip \Ex\left[Y_{t_{n+1}} \mid \bm{Y}_{n}=\bm{y}_{n}\right] & = p_{t_{n+1}} e^{\lambda_{t_n}(t_{n+1} - t_n)} \Ex\left[X_{t_n} \mid \bm{Y}_{n}=\bm{y}_{n}\right],
	\end{align*}
	where $\Ex\left[X_{t_n} \mid \bm{Y}_{n}=\bm{y}_{n}\right]$ is as given in  \cref{pro:Ex_Var}.
\end{proposition}